\documentclass[twocolumn,amssymb,nobibnotes,superscriptaddress,aps,prl]{revtex4-2}

\usepackage{graphicx}
\usepackage{color,amssymb}
\usepackage{hyperref}
\usepackage{amsmath}
\usepackage[dvipsnames]{xcolor}
\usepackage{tikz}

\usetikzlibrary{shapes.geometric}

\begin{document}

\title{Precise Twist Angle Determination in twisted WSe$_\mathbf{2}$ via Optical Moiré Phonons}%

\author{Nicolai-Leonid Bathen}
\affiliation{Institute of Physics, University of Münster, Münster, Germany}
\affiliation{Department of Materials Science and Engineering, Massachusetts Institute of Technology, Cambridge, MA, USA }
\author{Thorsten Deilmann}
\affiliation{Institute of Solid State Theory, University of Münster, Münster, Germany}
\author{Ana Senkić}
\affiliation{Institute of Physics, University of Münster, Münster, Germany}
\author{Hendrik Lambers}
\affiliation{Institute of Physics, University of Münster, Münster, Germany}
\author{Rami Dana}
\affiliation{Department of Materials Science and Engineering, Massachusetts Institute of Technology, Cambridge, MA, USA }
\author{Kenji Watanabe}
\affiliation{Research Center for Electronic and Optical Materials, National Institute for Material Science,1-1 Namiki, Tsukuba 305-0044,Japan}
\author{Takashi Taniguchi}
\affiliation{Research Center for Materials Nanoarchitectonics, National Institute for Material Science,1-1 Namiki, Tsukuba 305-0044,Japan}
\author{Frances M. Ross}
\affiliation{Department of Materials Science and Engineering, Massachusetts Institute of Technology, Cambridge, MA, USA }
\author{Julian Klein}
\affiliation{Department of Materials Science and Engineering, Massachusetts Institute of Technology, Cambridge, MA, USA }
\author{Ursula Wurstbauer}
\email{wurstbauer@uni-muenster.de}
\affiliation{Institute of Physics, University of Münster, Münster, Germany}
\affiliation{Center for Soft Nanoscience (SoN), University of Münster, Münster, Germany}

\date{\today}%
\begin{abstract}

Twisted bilayers of transition metal dichalcogenides (TMDC) form moiré superlattices resulting in moiré minibands in momentum space and hosting localized excitons in real space. While moiré superlattices provide access to Mott-Hubbard physics, their energy potential landscape and electronic correlations are highly sensitive to fluctuations of the twist angle, disorder and lattice reconstructions. However, fast and non-invasive experimental access to local twist angle and its spatial variations is challenging. Here, we systematically correlate twist angle variations of twisted WSe$_2$ bilayers across micrometer length scales using a combined lateral force microscopy (LFM) and a micro-Raman spectroscopy approach. These measurements uncover lateral variations in the twist angle by more than 1$^\circ$ across length scales relevant to optical and transport measurements. We demonstrate that twist angles in the range of $3^\circ < \alpha < 12^\circ$ show distinct Raman response from scattering on optical moiré phonons allowing twist angle determination with high precision and sub-micrometer spatial resolution under ambient conditions. These modes are  particularly sensitive in the low-angle twist regime, predicted to host emergent quantum phases. Our results establish micro-Raman spectroscopy of optical moiré phonons as a rapid, non-invasive probe to determine twist angle and  to screen local twist angle variations with a precision better than $\pm0.3^\circ$ and a lateral resolution below one micrometer. This methodology is also applicable to fully hBN-encapsulated heterostructures.  
\end{abstract}

\maketitle


Van der Waals (vdW) stacks represent a unique condensed matter platform to realize and control electronic correlation effects \cite{Kennes_review_21, Lado.2021, Sun.2024}, to study emerging quantum phases \cite{Mak_22} such as superconductivity \cite{10.1038/s41586-024-08116-2, 10.1038/s41586-024-08381-1, Xia_2025} and to host a variety of excitons ranging from localized excitons acting as single photon emitters to dense exciton ensembles featuring rich interaction physics \cite{Brotons_24}. 

Following the initial experimental demonstration of the Hofstadter butterfly in magnetotransport signals of graphene modulated by moiré superlattices formed by hexagonal boron nitride (hBN)-graphene bilayers \cite{Hofstadter.1976, Dean.2013}, correlated insulating, magnetically, excitonically or valley ordered, and (topological) superconducting states have been realized initially in half-metallic graphene-based moiré and moiré-free interfaces \cite{Lu.2019, Pantaleon_23, Balents_20, 10.1038/nature26160} as well as in homo- and heterobilayers of semiconducting group-VI transition metal dichalcogenides (TMDC) \cite{Brotons_24, Mak_22, Xiong_23, Wu_18}.

The potential landscape in these vdW bilayers is inextricably linked to the size of the moiré cell, which itself is determined by the twist angle and the difference in the lattice constants. In this way, localization in real space resulting from quasi-zero dimensional quantum mechanical confinement and the formation of flat moiré minibands in momentum space hosting emerging quantum phases crucially depend on the twist angle. Electron correlation phenomena arise  from an effective reduction in the kinetic energy of electronic states, enabled by moiré potentials, low-energy van Hove singularities, or band flattening achieved through carefully controlled interlayer coupling \cite{tang_simulation_2020, 10.1021/acs.nanolett.2c04169}. Interest in TMDC-based moiré platforms\cite{Mak_22, Sun.2024, 10.1038/s41586-021-03979-1, 10.1021/acs.nanolett.2c04169, 10.1103/physrevlett.121.026402} has been sparked by theoretical predictions of robust emergent quantum phases over a rather wide range of twist angles compared to "magic angle graphene bilayers" \cite{10.1038/s41467-021-24480-3}.\\

Unlike graphene, TMDC lattices are less rigid, making stacked and twisted TMDC homo- and heterobilayers prone to twist angle inhomogeneities and lattice reconstructions that are intrinsically highly periodic \cite{10.1038/s41563-020-00873-5,10.1038/s41567-021-01174-7, 10.1038/s41565-020-0682-9, 10.1088/2053-1583/acf5fb}. Spatial variations of moiré and reconstructed superlattice periodicities across nano- to micrometer length scales are knwon to modify energy potential landscape and mini-band formation, posing challenges for microscopic interpretation in optical far-field \cite{10.1103/physrevlett.133.046902, 10.1088/2053-1583/abf98e, 10.1002/adma.202008333} and transport studies \cite{An.2020, Rosenberger.2020, 10.1038/s41586-024-08116-2, 10.1038/s41586-024-08381-1}. This has been discussed as a possible reason why superconductivity in twisted TMDC bilayers has only recently been observed, and only for small nanoscale regions \cite{An.2020, Rosenberger.2020, 10.1038/s41586-024-08116-2, 10.1038/s41586-024-08381-1}. Knowledge of the local superlattice periodicity and hence twist angle is therefore crucial to understand the behavior of twisted bilayers and requires straightforward experimental access to local periodicity for each device with high precision. 
\begin{figure*}[ht!]
    \centering
     \includegraphics[width=1.0\linewidth]{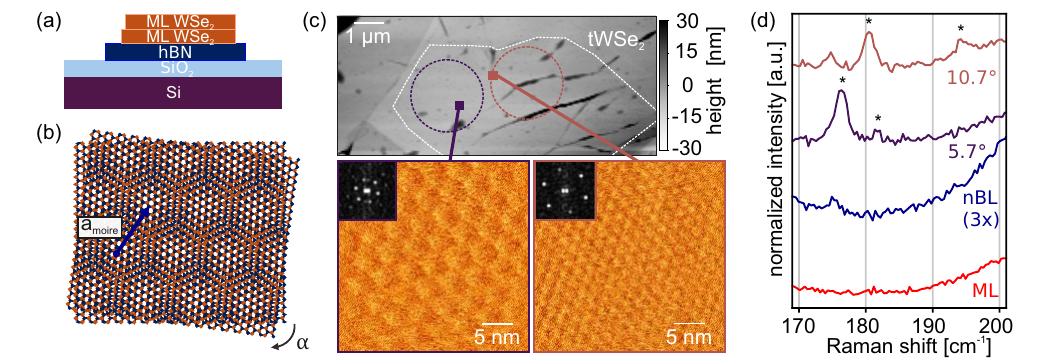}
    \caption{(a) Schematic showing (a) the layer sequence and (b) the moiré superlattice formed by a rigid tWSe$_2$ bilayer. (c) Top: Topographical atomic force microscopy (AFM) image of a representative tWSe$_2$ with the height information encoded in the grey scale. The tWSe$_2$ region is highlighted with a white dashed line. The dashed circles display regions at which the mean twist angle is determined from several LFM images. Bottom: Atomic-resolution LFM images collected at the marked locations showing moiré superlattice with varying $a_\mathrm{moire}$ of $5.7^\circ$ (left) and $10.7^\circ$ (right)  . Insets: Corresponding 2D FFTs. (d) Selected Raman spectra from monolayer WSe$_2$ (ML), natural bilayer WSe$_2$ (nBL) and from the two twisted regions marked by the dashed circles in (c). Modes marked by an asterisk (*) appear only for tWSe$_2$ bilayers. All measurements are performed in ambient at room temperature.}
  \label{fig:fig1_twist_disorder.pdf}
\end{figure*}
Global techniques such as controlled stacking (e.g. by the tear-and-stack method) or post-stacking characterization by second harmonic generation (SHG) measurements on non-stacked areas in a bilayer system are insufficient for the local determination of the twist angle due to local reconstructions and twist inhomogeneities \cite{SHG_twist_2019, SHG_2022}. Moreover, studies on the twist angle determination using SHG also showed lack of sensitivity in the twist angle range from $3^\circ$ to $7^\circ$, which is of particular interest for studying correlation physics \cite{10.1021/acs.nanolett.2c04169, 10.1038/s41563-020-0708-6, 10.1038/s41586-021-03815-6, 10.1038/s41586-024-08116-2, 10.1038/s41586-024-08381-1}. Nanoscopic techniques, such as scanning tunneling microscopy (STM) or scanning electron microscopy (SEM), are suitable for determining the moiré and reconstructed structure with high spatial precision, but they have several major disadvantages. STM is an experimentally slow and demanding technique that requires stable ultra-high vacuum conditions, and furthermore is only suitable for devices without the top hBN-encapsulation. SEM can only be employed for very thin top-hBN layers and the accelerated beam of electrons can affect the sample quality by defect creation and carbon contamination. It is therefore not ideal as a pre-chacterization method but is typically applied post-optical and/or transport investigation.  Micro-Raman spectroscopy is a powerful, non-invasive and easily-accessible technique to rapidly characterize TMDC layers \cite{Wurstbauer_2017}. Previous studies have identified Raman modes associated with acoustic moiré phonons whose appearance and energy depend on the moiré periodicity \cite{10.1002/adma.202008333, 10.1021/acsnano.8b05006, 10.1088/2053-1583/abf98e, dierke_moire_2025} and also with optical moiré phonons for twisted MoS$_2$ bilayers \cite{10.1021/acsnano.8b05006}. These studies show a lack of sensitivity in the twist angle range of interest for correlation physics ($3^\circ$ to $7^\circ$ \cite{10.1021/acs.nanolett.2c04169, 10.1038/s41563-020-0708-6, 10.1038/s41586-021-03815-6, 10.1038/s41586-024-08116-2, 10.1038/s41586-024-08381-1}). Furthermore, measurements of acoustic moiré phonons are experimentally demanding due to ultra-low phonon energies and additionally the precise determination of the absolute energy shift is more difficult than relying only on an energy difference as prominently shown e.g. for counting the layer number for MoS$_2$ determined by the energy difference of two optical modes \cite{Lee2010}.\\

Here, we demonstrate through a combined experimental and theoretical approach that optical moiré phonons in twisted WSe$_2$ (tWSe$_2$) bilayers serve as a distinctive fingerprint of the twist angle. These are conveniently probed by room-temperature, non-resonant Raman spectroscopy resulting in a fast and non-invasive approach suitable for both ambient and vacuum conditions. The typical energies of optical moiré phonons are above 150 cm$^{-1}$ making them easily accessible using standard dielectric laser blocking Raman filters, unlike acoustic moiré phonons that occur well below 80 cm$^{-1}$, an energy range only accessible with specialized experimental techniques. By combining micro-Raman spectroscopy and lateral force microscopy (LFM) measurements on identical samples, we show that spatial twist angle variations can be quantified with high precision. This combined approach demonstrates that micro-Raman spectroscopy is suitable to determine the local twist angle $\alpha$ with a precision better than $\pm0.3^\circ$ in an extended range covering $3^\circ < \alpha < 12^\circ$. The precise determination of the twist angle can also be achieved in fully hBN encapsulated tWSe$_2$ bilayers and we expect the approach to be generally suitable for field-effect structures with semi-transparent top-gate electrodes.  \\


In this work, we study seven tWSe$_2$ bilayer samples with twisted areas spanning several to tens of micrometer laterally and  that include a wide range of twist angles $0^\circ <$ t $< 12^\circ$. The tWSe$_2$ bilayers are stacked on top of a multilayer hBN supported on a Si/SiO$_{2}$ substrate, as schematically shown in \autoref{fig:fig1_twist_disorder.pdf}(a). The samples are prepared by micromechanical cleavage and viscoelastic dry transfer including an annealing step as described in detail in the Experimental section. By applying a rotation between the two WSe$_{2}$ monolayers during the transfer process, a superlattice is created with moiré lattice constant $a_\mathrm{moire}$ that is connected to the twist angle $\alpha = 2\arcsin(\frac{a_\mathrm{WSe2}}{2a_\mathrm{moire}})$ between the two layers. A schematic representation of the expected moiré superlattice pattern for a rigid TMDC bilayer is illustrated in \autoref{fig:fig1_twist_disorder.pdf}(b). We collect extensive LFM measurements over large areas on tWSe$_2$ bilayers to determine statistically and quantitatively the moiré superlattice structure in real space with high lateral resolution \cite{10.1103/physrevb.78.045432, 10.1007/s11249-010-9677-2, 10.1103/physrevb.81.155412, 10.1021/acs.langmuir.3c01546}. We conduct AFM measurements to obtain a topographical overview of the samples allowing us to avoid bubble, fold and scratch regions. Then, we conduct high-resolution LFM measurements and, finally record micro-Raman maps on the identical samples and under comparable conditions. In \autoref{fig:fig1_twist_disorder.pdf}(c), an AFM topography image with the height shown in grey scale is displayed with the tWSe$_2$ bilayer region indicated by white dashed lines. Two representative high-resolution LFM measurements recorded at positions about one micrometer apart are highlighted in the topography image by colored square markers. The LFM images reveal the atomic lattice and an additional hexagonal periodicity, which is interpreted as the moiré superlattice periodicity. The moiré lattice constants $a_\mathrm{moire}$ change substantially between the two positions on the flake. To quantify $a_\mathrm{moire}$ and its lateral changes, we analyze the LFM images by 2D Fast Fourier Transformation (2D FFT) as included in each LFM image. The 2D FFT clearly shows the hexagonal structure of both the underlying WSe$_2$ lattice (see Supplemental Information), with lattice constant $a_\mathrm{WSe2}$, and the moiré superlattice $a_\mathrm{moire}$. The 2D FFT allows the direct extraction of the periodicity of the reciprocal lattices and therefore the determination of the real space values $a_\mathrm{moire}= 3.4\,\mathrm{nm}$ and $a_\mathrm{moire}= 1.8\,\mathrm{nm}$ corresponding to twist angles of $5.6^\circ$ and $10.5^\circ$, respectively.

To ensure robust statistics, we calculate mean values of the twist angle deduced by averaging over the 2D FFT results from $\geq$ 6 LFM images taken within a circular region with a diameter between 1.6 and 2.4 µm. This area is chosen to relate to the typical spot sizes in optical measurements in cryogenic environment [c.f. dashed circles in \autoref{fig:fig1_twist_disorder.pdf}(c)]. The procedure is explained in more detail in the Supplemental Information. For the tWSe$_2$ bilayer region shown in the AFM topography in \autoref{fig:fig1_twist_disorder.pdf}(c), the mean twist changes by several degrees from $(5.71 \pm 0.07)^\circ$ to $(10.65 \pm 0.04)^\circ$. The drastic change in the moiré lattice over few micrometer-scale distances is in agreement with recent reports in the literature \cite{10.1038/s41563-020-00873-5,10.1038/s41567-021-01174-7,10.1038/s41565-020-0682-9,10.1088/2053-1583/acf5fb}.

In \autoref{fig:fig1_twist_disorder.pdf}(d), micro-Raman spectra are compared between monolayer (ML), natural bilayer (nBL) and the two tWSe$_2$ spectra taken on the positions marked with dashed circles in panel (c). The Raman spectra are recorded in ambient conditions using an excitation laser of 532\,nm with diffraction limited spot size. The spectral range is centered around the optical $E_g^2$-mode (174 cm$^{-1}$), which is absent in the ML WSe$_2$ spectrum, as expected since it is Raman inactive under the employed back-scattering geometry \cite{10.1103/physrevb.88.195313}. The $E_g^2$-mode appears for the natural bilayer as well as for tWSe$_2$ bilayers and does not shift with twist angle. Two additional modes at the high energy-side, marked by an asterisk ($^\ast$) in \autoref{fig:fig1_twist_disorder.pdf}(d), appear only for the tWSe$_2$ bilayers. Their energies depend non-trivially on the twist angle and are interpreted as moiré phonons \cite{10.1002/adma.202008333, 10.1021/acsnano.8b05006, 10.1088/2053-1583/abf98e} related to the $E_g^2$ mode. These modes are Raman-active due to the superimposed moiré superlattice, which gives rise to the formation of mini-Brillouin zones in momentum space causing back-folding of the phonon dispersion by the reciprocal moiré superlattice vector $\vec{q}_\mathrm{moire}$ \cite{10.1103/physrevlett.37.1145,10.1016/0038-1098(82)90319-2, 10.1103/physrevlett.54.2111, 10.1063/1.1691171}. The non-trivial dependence of these optical moiré phonon modes on the twist angle is explained below.

\begin{figure}[ht!]
    \centering
    \includegraphics{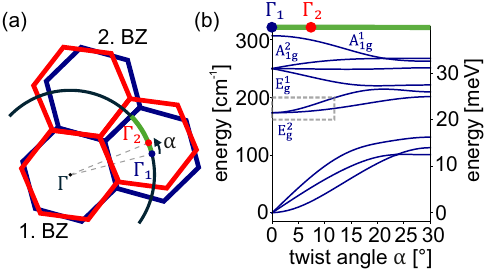}
    \caption{(a) Schematic of the first and second Brillouin zone of a tWSe$_2$ bilayer. The $\Gamma$-points of the second Brillouin zones $\Gamma_1$ and $\Gamma_2$ are displaced along the curved trajectory (green line). Their connecting vector $\vec{q}_{moir\acute{\text{e}}}$ denfines the size of the mini Brillouin zone that depends on the twist angle $\alpha$. 
    (b) DFT calculated phonon dispersion for a monolayer WSe$_{2}$. According to (a), the momentum axis coordinate can be translated into the twist angle $\alpha$ ($x$-axis) for zero-momentum phonons in the mini-Brillouin zone. Dashed rectangle denotes the optical moiré phonons studied in this work experimentally by Raman spectroscopy.}
    \label{fig:fig2_theory.pdf}
\end{figure}

Theoretical first-principles investigations of moiré phonons for such small twist angles $\alpha$ are challenging. The moiré cells given by $a_\mathrm{moire}$ become huge in real space and exact calculations are not feasible. To provide a reasonable description of the optical moiré phonon modes, a straightforward approach is to neglect the (often) small interaction between the layers \cite{10.1002/adma.202008333, 10.1021/acsnano.8b05006} and to use the phonon dispersion for a WSe$_{2}$ monolayer. By rotating two mono-layers relative to each other, their reciprocal space is also rotated as indicated in the schematic shown in \autoref{fig:fig2_theory.pdf}(a). Thus, the positions of the $\Gamma$ points of the two individual layers are misaligned in the second Brillouin zones.
The final momentum mismatch between the two $\Gamma$-points corresponds to $\vec{q}_\mathrm{moire}$ such that the momentum-axis in the dispersion of the underlying WSe$_{2}$ monolayer can be translated into the twist angle $\alpha$ of the twisted bilayer as sketched by the green curved line connecting the two $\Gamma$ points in \autoref{fig:fig2_theory.pdf}(a).
In this way, the phonon dispersion displayed in panel \autoref{fig:fig2_theory.pdf} (b) can be represented as a function of twist angle $\alpha$ for zero-momentum phonons in the mini-Brillouin zone for twisted bilayers. The positions of the two $\Gamma$-points are included schematically on the top axis. We note that this approach neglects the effects of interactions between the layers, such as atomic reconstructions \cite{Woźniak.2020, Enaldiev.2020, Carr.2018,  Maity.2021}. The phonon dispersion employed for the WSe$_2$ monolayer has been calculated by density functional theory (DFT) \cite{GdWTMDC, Mann_2024}.\\
In the described representation all phonon branches should be understood as zero-momentum in the mini-Brillouin zone. As a direct consequence momentum conservation in Raman experiments using visible light is fulfilled. This implies that the energy and number of optical moiré phonon modes in non-resonant Raman spectra crucially depend on the twist angle $\alpha$ and serve as a sensitive probe. In particular, Raman spectra of tWSe$_2$ bilayers should display in addition to the $\Gamma$-point $E_g^2$-mode (corresponding to $\alpha=0^\circ$ in this representation and called E$_{0}^{\Gamma}$ in the following) two additional slightly blue-shifted Raman modes with energies depending on the twist angle. The modes marked with an $^\ast$ in the Raman spectra displayed in \autoref{fig:fig1_twist_disorder.pdf}(d) are potential candidates for those optical moiré phonon modes, which we denote as E$_{1, 2}^{t}$ .

\begin{figure}[ht!]
    \centering
    \includegraphics{fig3_moire_phonons.pdf}
    \caption{(a) Selected Raman spectra for different mean twist angles in the spectral range of the $E_g^2$-mode, showing the presence of three different modes with the energy of E$_{0}^{\Gamma}$ being independent of the twist angle and the energy of optical moiré phonons E$_{1, 2}^{t}$ highly dependent on $\alpha$. The Raman spectra are normalized to the intensity of the mode at 137\,cm$^{-1}$. For clarity, spectra are shifted vertically by a constant value. (b-d) Selected representative LFM images from the areas in which the spectra in (a) are collected. The moiré cell sizes are clearly visible in real-space image as well as in 2D FFTs.}
    \label{fig:fig3_moire_phonons.pdf}
\end{figure}

Following these theoretical considerations, we demonstrate that the expected optical moiré phonons  E$_{1,2}^{t}$ related to the E$_g^2$-branch are indeed prominently observable in Raman experiments. To this end, we record spatial micro-Raman maps in ambient conditions with an excitation wavelength of $\lambda_\mathrm{Raman}=532$nm on the tWSe$_{2}$ regions with a high-resolution spectrum (energy resolution better than 0.1cm$^{-1}$) for each pixel. Details on the Raman maps are summarized in the Experimental section and Supplemental Information.

Thorough analysis of the Raman spectra verifies that occurrence and energy of the highlighted optical moiré phonon modes E$_{1,2}^{t}$ are indeed suitable to determine the twist angle with high precision in a rather large twist angle range of at least $3^\circ < \alpha < 12^\circ$. \autoref{fig:fig3_moire_phonons.pdf}(a) demonstrates this precision by comparing Raman spectra collected from six different tWSe$_2$ samples with various twist angles $0.3^\circ\leq\alpha\leq11.5^\circ$ with the mean twist angles deduced from LFM measurements. The uncertainty of the mean twist angle from LFM is below $\pm 0.1^\circ$ as detailed in the Supplemental Information. Representative LFM maps are shown in \autoref{fig:fig3_moire_phonons.pdf}(b-d) for a mean twist angle of about $3.7^\circ$, $6.7^\circ$ and $11.5^\circ$, respectively. The changes in the moiré lattice periodicity and evidence for macroscopic reconstructions indicated by triangular or hexagonal superimposed periodic pattern are visible. The 2D FFT proves the moiré periodicity enabling quantitative analysis of the twist angle. To achieve better spatial correspondence between the high-resolution LFM measurements and the Raman maps, we averaged Raman spectra (20-60 depending on the data set) from a circular region similar to the one from which the mean twist angle was determined (dashed circles in  \autoref{fig:fig1_twist_disorder.pdf}(c) ). Comparison between the point Raman spectrum with a lateral resolution of about 400nm and the spatially averaged one extracted from the map is shown in the Supplemental Information. The averaged Raman spectra are then normalized to the mode at 137 cm$^{-1}$. Three distinct modes appear in those Raman spectra for twist angles larger than $3^\circ$. For the whole twist angle range, we find one mode (174 cm$^{-1}$) that is independent of the twist angle and interpreted as the zone center optical phonon $E_{g}^2$, known from the natural bilayer and labeled E$_{0}^{\Gamma}$ in \autoref{fig:fig3_moire_phonons.pdf}(a). This mode serves as a sample- and instrument-independent reference point to measure the energy difference $\Delta E_{1,2}$ between the zone center E$_{0}^{\Gamma}$ mode and the twist-sensitive optical moiré phonon modes labeled $E_1^t$ and $E_2^t$, respectively. The energies of two optical moiré phonon modes both shift to higher energies with increasing twist angle but in different ways. For twist angles smaller than $3^\circ$, these two modes can be clearly resolved in a careful line-shape analysis of the recorded spectra using a sum of Lorentzian functions. Since the modes have greater overlap for smaller twist angles $\alpha$, a quantitative analysis is challenging and comes with a higher uncertainty in the determination of the twist angle. Nevertheless, we emphasize that even for a twist of only $0.3^\circ$ an asymmetric lineshape is observable, still allowing a qualitative analysis in this low twist angle regime for $\alpha < 3^\circ$.

\begin{figure}[ht!]
    \centering
    \includegraphics{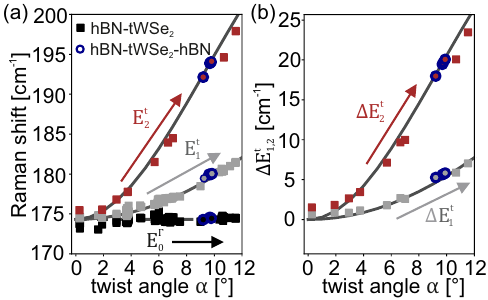}
    \caption{Correlating twist angle $\alpha$ and energies of optical moiré phonons in tWSe$_2$ samples. Filled squares represent hBN-tWSe$_2$ and blue outlined symbols fully encapsulated hBN-tWSe$_2$-hBN samples. (a) Energies of $E_0^\Gamma$ ($E_g^2$) and the two optical moiré phonons $E_1^t$ and $E_2^t$.  Solid grey lines represent DFT calculated dispersion. For the hBN-tWSe$_2$ samples, the twist angle is determined from LFM, while for the fully encapsulated hBN-tWSe$_2$-hBN samples, we used the theoretical model. (b) Energy difference  $\Delta E_1^t$ and $\Delta E_2^t$ of optical moiré phonons relative to $E_0^\Gamma$ for varying twist angle.}
    \label{fig:fig4_moire_phonons_comparison.pdf}
\end{figure}

Our findings are robust and universal for all samples we have investigated by comprehensive LFM and Raman mapping experiments following the protocol described above. In \autoref{fig:fig4_moire_phonons_comparison.pdf}(a), we compare extracted Raman mode energies for the E$_{0}^{\Gamma}$, $E_1^t$ and $E_2^t$ from all seven hBN-WSe$_2$ samples and for various areas with the calculated moiré-phonon dispersion. We find excellent agreement between theory and experiment. This agreement substantiates the applicability of the simplified theoretical approach utilizing zone-backfolding and neglecting interlayer coupling. The E$_{0}^{\Gamma}$-mode is unchanged by the twist angle as it is the zero momentum $E_g^2$ mode at the zone center of the underlying WSe$_{2}$ lattice, while the $E_1^t$ and $E_2^t$ mode energies follow the expected dispersion in dependence of the twist angle. As a direct consequence, the energy differences $\Delta E_{1}= E_1^t-E_{0}^{\Gamma}$ and $\Delta E_{2}= E_2^t -E_{0}^{\Gamma}$ between the zone-center and the moiré phonons in combination serve as a robust and precise figure of merit for the twist angle, independent of the sample, small perturbations or small uncertainties in the instrumental alignment of the Raman spectrometer. The energy differences $\Delta E_{1,2}$ of the experimentally observed modes closely follow the calculated dispersion as summarized in \autoref{fig:fig4_moire_phonons_comparison.pdf}(b), confirming that the expected values for $\Delta E_{1}$ and $\Delta E_{2}$ at a certain twist angle are in good agreement. Most importantly, the agreement is excellent in the critical range for correlation physics between 3$^\circ$ and 7$^\circ$, a range where previous studies failed to demonstrate a precise determination of the twist angle \cite{10.1002/adma.202008333, 10.1021/acsnano.8b05006, 10.1088/2053-1583/abf98e}.\\

So far, only single-sided hBN interfaced tWSe$_2$ (hBN-tWSe$_2$) structures have been discussed, since LFM investigation does not allow for full hBN encapsulation. As a proof of concept, we prepared fully hBN encapsulated tWSe$_2$ (hBN - tWSe$_2$ - hBN) samples and performed micro-Raman measurements as described above and displayed in the Supplemental Information. For the fully encapsulated samples, the same three modes E$_{0}^{\Gamma}$, E$_{1}^{t}$ and E$_{2}^{t}$ belonging to the $E_g^2$  phonon branch show up in our Raman spectra (see Supplemental Information). The twist angle is determined from comparison of Raman mode energies determined from a combined twofold Lorentzian fit and comparison of $\Delta E^t_{1}$ and $\Delta E^t_{2}$ with the theoretical prediction. The extracted mode energies for $E_0^\Gamma$,  $E_1^t$ and $E_2^t$  and the resulting differences $\Delta E_{1}$ and $\Delta E_{2}$ are included in \autoref{fig:fig4_moire_phonons_comparison.pdf}(a,b), respectively. 

Both $\Delta E^t_{1}$ and $\Delta E^t_{2}$ suggest a twist angle of about $10^\circ$. This value is in the twist range roughly expected from the angular alignment between the two individual WSe$_{2}$ monolayers during the stacking process. The spectral analysis of the zone-center $E_0^\Gamma$  and the twist dependent moiré phonons $E_1^t$ and $E_2^t$ of the optical $E_g^2$ phonon branch allows the rather precise determination of the local twist angle in tWSe$_{2}$ with diffraction limited spatial resolution on fully encapsulated samples. By such Raman investigations the absolute twist angle including lateral changes and twist disorder can be resolved as shown in \autoref{fig:fig1_twist_disorder.pdf}. Moreover, we expect that our approach is universal and the optical moiré phonons can also be used to determine the twist angle in twisted bilayers prepared from other group-VI TMDC monolayers and heterobilayers. However, one should have in mind that the method is sensitive to the moiré superlattice constant $a_{moir\acute{e}}$ and only indirectly to the twist angle $\alpha$ given by the zone-backfolding approach. Hence, the real space lattice constants need to be considered in correlating $a_{moir\acute{e}}$ with the twist angle. 

To conclude, we demonstrate that optical moiré phonons associated with the optical $E_g^2$ phonon branch are active in non-resonant Raman scattering and suitable for determining the local twist angle. In this way, the twist angle in tWSe$_2$ can be determined over a large range of angles  $3^\circ < \alpha < 12^\circ$ with a precision better than $0.3^\circ$. The method is rapid, non-destructive and can be easily performed at room-temperature at atmospheric conditions with diffraction limited resolution of better than a micrometer using standard Raman microscopes with standard dielectric filters. The relationship of Raman spectra and local twist angles was substantiated by an intensive investigation of a total of seven tWSe$_2$ bilayers using LFM and micro-Raman experiments. By a proof-of-concept study, we confirmed that this approach to determine the twist angle is equally applicable to fully hBN-encapsulated structures. 
Since the analysis consists  of the energy difference $\Delta E_{1}$ and $\Delta E_{2}$ between the two observed moiré phonon modes with the twist-independent zone-center mode $E_0^\Gamma$, micro-Raman mapping allows not only the precise determination of the twist angle, but also the investigation of the lateral twist angle variations, both within individual samples and across different samples with decent precision. Lastly, the method is versatile across multiple experimental setups. The excellent agreement between experimental data and theoretical predictions highlights the effectiveness of the simple back-folding model, supporting its broader applicability to other TMDC-based artificial bilayers and beyond even at small twist angles. Due to the well-known issue with lateral twist disorder with twist changes of more than one degree within micrometer length-scale \cite{Rosenberger.2020, 10.1021/acs.nanolett.2c03676, 10.1038/s41563-023-01595-0, 10.1038/s41567-021-01174-7}, the precise knowledge of both the local twist angle and of twist inhomogeneity across a sample is of great importance for optical and transport studies in the rich field of twistronics spanning from single exciton to correlated exciton ensembles \cite{10.1038/s41586-019-0957-1, Xiong_23, Tran.2019} and emerging quantum phases in the context of Mott-Hubbard model physics \cite{10.1038/s41586-021-03979-1, 10.1021/acs.nanolett.2c04169, 10.1103/physrevlett.121.026402}.

\section{Methods}
\textbf{Sample preparation}
Individual flakes, mechanically exfoliated from bulk crystals (WSe$_2$: hq graphene, hBN: hq graphene for fully encapsulated samples and \cite{10.1038/nmat1134} for single-sided encapsulated samples), were assembled using the poly-carbonate (PC) method and then transferred to a Si/SiO$_2$ substrate. To achieve clean surfaces for the LFM measurements, the samples were soaked in chloroform for at least 20\,hours, and then vacuum annealed above 180\,°C for several hours. To ensure a high Raman intensities, based on the light propagation effects \cite{10.1007/s12274-014-0424-0}. a 90\,nm Si/SiO$_2$ substrate with thin hBN (1.2-11\,nm) was targeted.

\textbf{Raman spectroscopy and Raman mapping:}
 Raman spectra were measured with 532\,nm excitation wavelength at room temperature in ambient. For the measurements of single-sided hBN samples (hBN-tWSe$_2$) a WITec alpha300 system (Zeiss LD EC Epiplan-Neofluar HD Dic 100x / 0.75) was used with a power of 0.5mW and an integration time per spot up to 20 s. For the measurements on fully encapsulated hBN sample, a Horiba Scientific, LabRAM HR evolution system (Carl Zeiss Microscopy 100x / 0.9) was utilized with a power of 0.2 mW and an integration time of 30s. Raman maps were recorded with a step size of 300 nm. All data were shifted to ensure that the Si peak \cite{10.1016/0038-1098(72)91127-1, 10.1149/2.0061509jss} is positioned at 520\,cm$^{-1}$, ensuring consistent calibration when comparing data sets. For all measurement a grating with 1800 mm$^{-1}$ constant was used.

\textbf{Lateral Force Microscopy:}
All samples, except the one with the nominal 0.3$^\circ$ twist angle, were studied with LFM with image size ranging from 30x30 nm to 80x80 nm. In the 0.3$^\circ$ twisted case, because its moiré lattice constant was $>20$ nm, LFM was not suitable, so we used piezoresponse force microscopy (PFM) to determine the twist angle \cite{McGilly2020} in images of size $700 \times 700\,\text{nm}^2$. All scanning probe measurements were performed with an Asylum Research Cypher VRS AFM instrument with ElectriMulti75-G tips from BudgetSensors with following properties: spring constant: 3\,N/m (1-7\,N/m), resonance frequency: 75\,kHz (60-90\,kHz), length 225\,µm (215-235\,µm)]. 

\textbf{Acknowledgement}
The authors gratefully acknowledge financial support by the German Science Foundation (DFG) via Grants No.443274199 and 556436549 (WU 637/7-2,8-1), No. 426726249 (DE 2749/2-1 and DE 2749/2-2) and the priority program 2244 (2DMP). J.K. and F.M.R. acknowledge funding through the NSF Trailblazer Award Number 2421694 and DOE-BES Award Number DE-SC0025387.This work was carried out in part through the use of MIT.nano's facilities. N.L.B., A.S. and U.W. acknowledge project from Helmholtz-Institut Münster "EFoBatt" Grant No. 13XP5129 for using commercial Horiba Scientific, LabRAM HR evolution commercial Raman microscope. K.W. and T.T. acknowledge support from the JSPS KAKENHI (Grant Numbers 21H05233 and 23H02052), the CREST (JPMJCR24A5), JST and World Premier International Research Center Initiative (WPI), MEXT, Japan. The authors gratefully acknowledge the Gauss Centre for Supercomputing e.V. (www.gauss-centre.eu) for funding this project by providing computing time through the John von Neumann Institute for Computing (NIC) on the GCS Supercomputer JUWELS \cite{JUWELS} at Jülich Supercomputing Centre (JSC).



%

\clearpage
\pagebreak

\widetext
\begin{center}
\textbf{\large Supplemental Information:\\
Precise Twist Angle Determination in twisted WSe$_\mathbf{2}$ via Optical Moiré Phonons}
\end{center}

\setcounter{equation}{0}
\setcounter{figure}{0}
\setcounter{table}{0}
\setcounter{page}{1}
\makeatletter
\renewcommand{\theequation}{S\arabic{equation}}
\renewcommand{\thefigure}{S\arabic{figure}}
\renewcommand{\bibnumfmt}[1]{[S#1]}
\renewcommand{\citenumfont}[1]{S#1}

\subsection*{LFM image and Raman spectroscopy data analysis}

\begin{figure*}[ht!]
    \centering
    \includegraphics[width=1.0\linewidth]{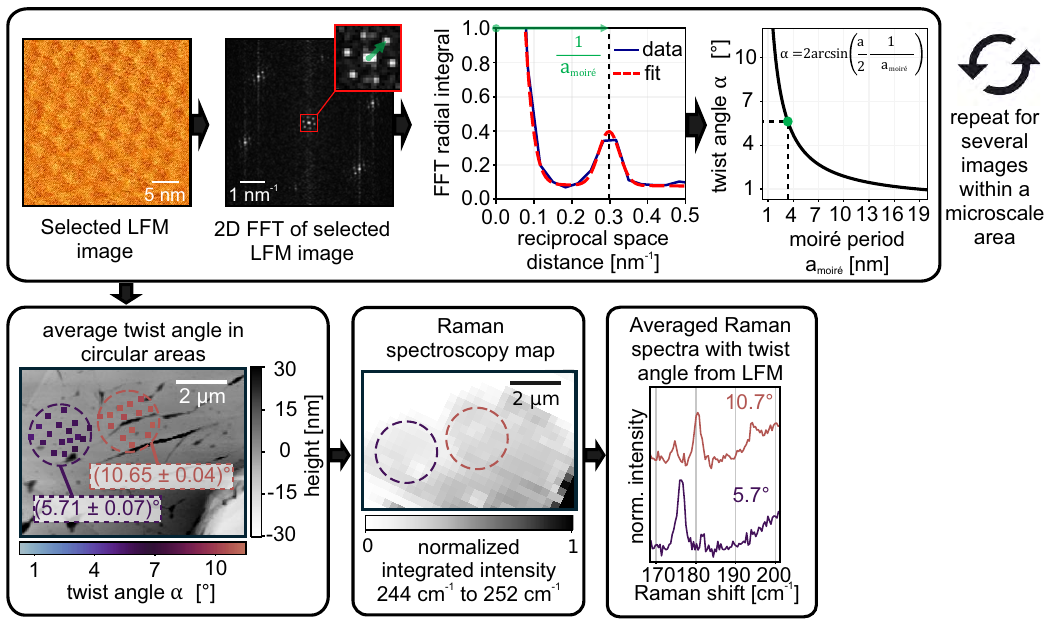}
    \caption{Workflow used to determine the average twist angle at different positions on the sample using LFM, and to extract averaged spectra from a Raman map at the same positions.}
    \label{fig:sup_minus1}
\end{figure*}

The data analysis process used to determine the average twist angle at different positions on the sample using lateral force microscopy (LFM), and to extract averaged spectra from a Raman map at the same positions is sketched in \autoref{fig:sup_minus1} and explained in the following. LFM was used to measure the local moiré lattice in nanoscale images (30x30\,nm to 80x80\,nm). The moiré periodicity $a_\mathrm{moire}$ was determined with a radial integral of the 2D Fast Fourier Transformation (FFT). From that, the twist angle $\alpha$ was calculated by 

\begin{equation}
\alpha = 2 \arcsin \left( \frac{a}{2} \frac{1}{a_{\text{moiré}}} \right)
\label{eq:moire}
\end{equation}

Here $a$ is the lattice constant of WSe$_2$, which was set to 0.328\,nm \cite{Hulliger.1976}. The process was repeated at least 6 times within a circular microscale area to obtain reasonable statistics for calculating the mean twist angle in this area. The area corresponds to a typical laser spot size in the cryogenic conditions.

For the sample with a nominal twist angle of 0.3° piezoresponse force microscopy (PFM) was used because it was found to be advantageous for images larger than 200x200\,nm and small twist angles close to 0\,°. The mean twist angle and the size of the single moiré cells was determined by tracing the borders of the moiré cells with circles and triangles in the open-source image processing software \textit{fiji} \cite{Schindelin2012}. From the determined area $A$ the moiré periodicity $a_\mathrm{moire}$ was estimated by the diameter of the circle $ d_\circ = 2 \sqrt{\frac{A}{\pi}}$ and the height of the triangle $d_\triangle =  \sqrt{A \cdot \sqrt{3}}$ where triangles are approximated to be equilateral. Since in this twist-angle regime, large changes in the moiré cell size on the order of 10 nm lead only to small changes in the corresponding twist angle on the order of 0.1°, this approach enabled similarly precise twist-angle determination.

Raman maps were recorded on the same samples with a step size of 300\,nm, 0.5\,mW excitation power and with an integration time per spot $\leq 20\,$s. For the fully encapsulated sample, step size was 300\,nm, 0.2\,mW excitation power and the integration time per spot was 3x10s. Averaged spectra are extracted from the same microscale areas investigated in the LFM data. The resulting spectra are directly linked to the mean twist angle determined from the corresponding region in LFM. Subsequently, the averaged spectra underwent a Lorentzian fitting processes. 


\subsection*{Additional Raman spectra}

\begin{figure*}[ht!]
    \centering
    \includegraphics[width=0.9\linewidth]{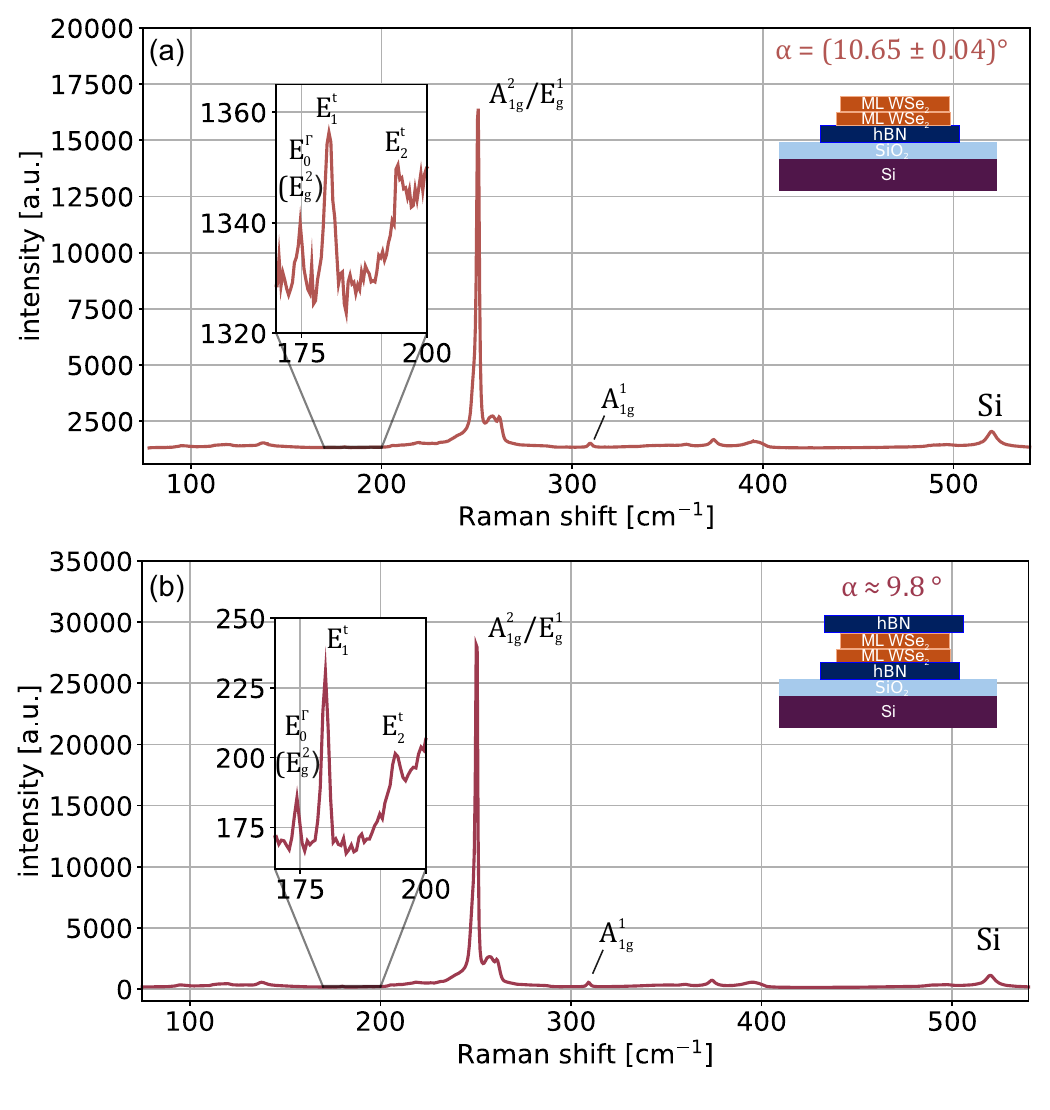}
    \caption{Representative Raman spectrum for (a) tWSe$_2$ with bottom hBN and (b) fully encapsulated tWSe$_2$, both on a Si/SiO$_2$ substrate, measured with 532\,nm excitation wavelength. The mean twist angle was determined by (a) LFM and (b) optical moiré phonons. Marked are the first-order Raman modes of WSe$_2$ by their symmetry, the moiré phonons and the visible Si Raman mode based on Refs. \cite{Blaga.2024,10.1088/2053-1583/aa77d4}. }
    \label{fig:sup0}
\end{figure*}

\begin{figure}[ht!]
    \centering
   \includegraphics[width=1\linewidth]{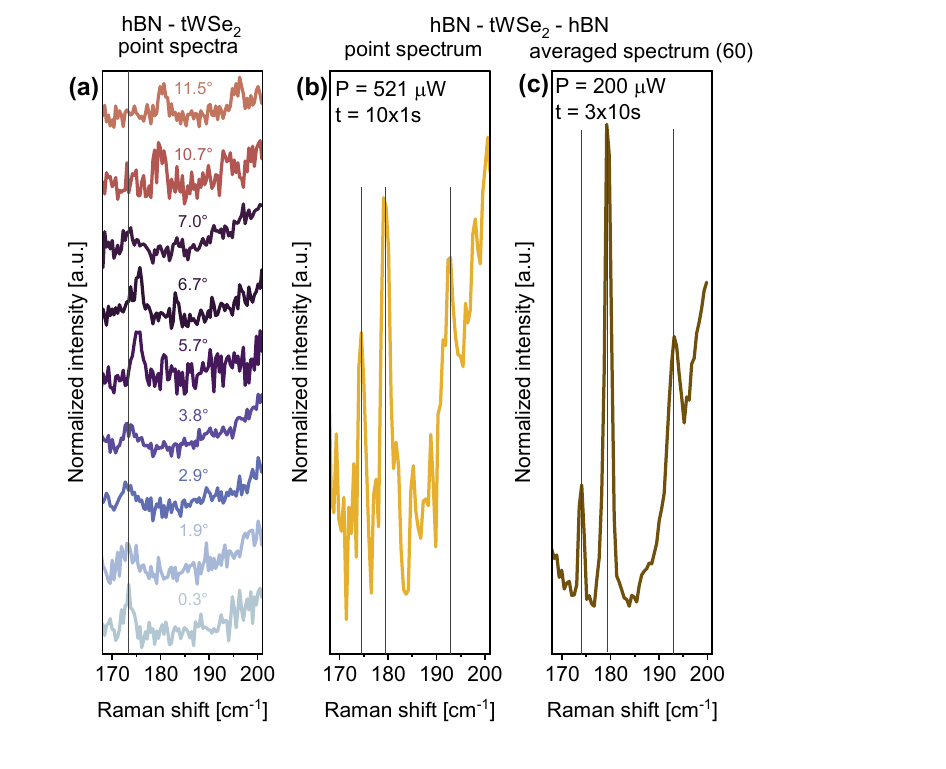}
   \caption{(a) Raman spectra taken on individual points on single-sided hBN interfaced tWSe$_2$ with diffraction limited spatial resolution from the areas used to obtain the averaged spectra displayed in main figure 3 are taken from. Spectra are taken in ambient conditions with $\lambda_\mathrm{Raman}=532$nm, P = 0.5mW and integration time of 20s or less.
   (b) and (c) Raman spectra taken on fully hBN encapuslated tWSe$_2$ under similar conditions. Comparison of Raman spectra taken on an individual diffraction limited point [P = 521µW and integration time of 10s] (b) with an spatially averaged spectrum (c) following the procedure described in the main text by  averaging over an area of about 1.6 µm diameter within an homogeneous part on the sample. The comparison displays good agreement with superior signal to noise ration for the averaged spectrum. Large-area Raman maps with high spatial resolution allows for detection of twist angle inhomogeneities within a reasonable measurement time. The spatially averaged Raman spectra over a homogeneous twist area is in good agreement with the point spectrum taken with longer acquisition time in the same area, justifying the averaging process.}
   \label{fig:sup4}
\end{figure}
\clearpage

\subsection*{Additional LFM and Raman spectroscopy data comparison}

\begin{figure*}[ht!]
    \centering
    \includegraphics[width=0.75\linewidth]{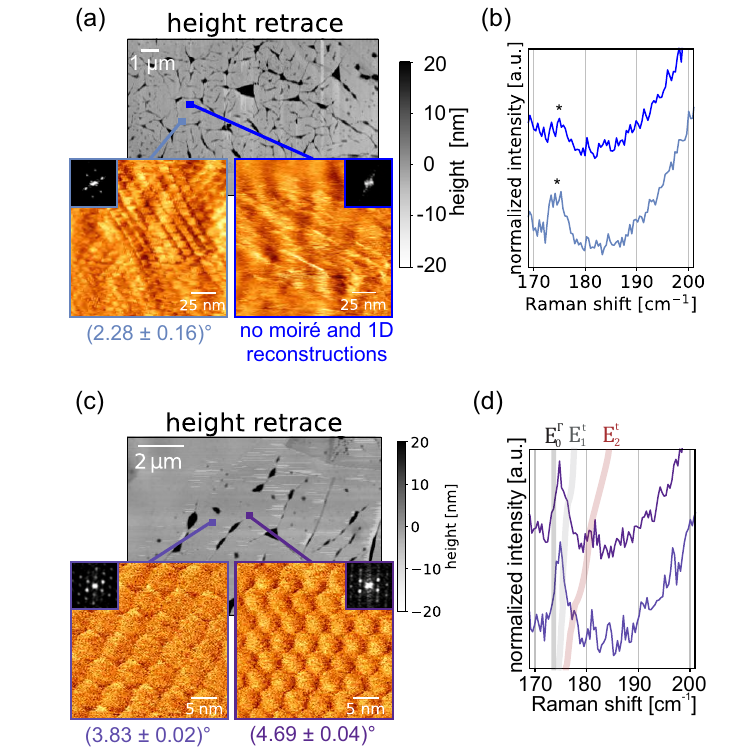}
    \caption{(a,c)  The height traces of two studied samples with relocated LFM scans, which show the moiré lattice at two distinct locations on the sample. (b,d) Selected Raman spectra from the regions marked in (a,c).}
    \label{fig:sup1}
\end{figure*}
\clearpage

\subsection*{Values and statistical uncertainties for twist angles of Raman data shown in Figure 3}
\begin{table}[h]
    \centering
\begin{tabular}{||c || c | c | c | c | c | c | c | c | c ||}
 \hline
 Mean value ($^\circ$) &  0.26 & 1.90 & 2.93 & 3.72 & 5.71 & 6.71 & 6.97 & 10.65 & 11.51 \\ [1ex] 
 \hline\hline
 Uncertainty ($^\circ$) & 0.01 & 0.09 & 0.04 & 0.02 & 0.07 & 0.03 & 0.03 & 0.04 & 0.04 \\ [1ex] 
 \hline
\end{tabular}
    \caption{Mean values of the twist angles and corresponding uncertainties determined from LFM measurements, for six different tWSe$_2$ samples. Raman signatures of optical moiré phonons $E^t_{1,2}$ stemming from these twist angles are shown in Figure 3 in the main text.}
    \label{twist_table}
\end{table}


%

\end{document}